\documentclass[journal]{IEEEtran}

\usepackage{graphicx}
\usepackage[dvipsnames]{xcolor}
\usepackage{amssymb}
\usepackage{amsmath}
\usepackage{optidef}
\usepackage{array}
\usepackage{algpseudocode}
\usepackage{multirow}
\usepackage{algorithm}
\usepackage{algcompatible}
\usepackage{cite}
\usepackage{csquotes}
\usepackage{float}
\usepackage{todonotes}
\usepackage{soul}
\usepackage{diagbox}
\usepackage[bottom]{footmisc}
\usepackage{lipsum}
\usepackage{enumerate}
\usepackage{url}
\usepackage[subnum]{cases}
\usepackage{optidef}
\usepackage[normalem]{ulem}

\DeclareMathSizes{10}{9}{8}{7}
\usepackage[font={small}]{caption, subfig}
\usepackage{mathtools,amssymb,lipsum}
\newcommand\windowhyphen{\operatorname{W-MSA}}
\newcommand\shiftwindowhyphen{\operatorname{SW-MSA}}
\ifCLASSINFOpdf
\else
\fi
\begin{document}
%
\title{\huge Semantic Communication Enabled 6G-NTN Framework: A Novel Denoising and Gateway Hop Integration Mechanism}
%
%
%

\author{Loc X. Nguyen, Sheikh Salman Hassan,~\IEEEmembership{Member,~IEEE,}  Yan Kyaw Tun,~\IEEEmembership{Member,~IEEE,} \\ Kitae Kim,~\IEEEmembership{Member,~IEEE,} Zhu Han,~\IEEEmembership{Fellow,~IEEE}, and Choong Seon Hong,~\IEEEmembership{Fellow,~IEEE}

\thanks{Loc X. Nguyen, Sheikh Salman Hassan, Kitae Kim, and Choong Seon Hong are with the Department of Computer Science and Engineering, Kyung Hee University, Yongin-si, Gyeonggi-do 17104, Rep. of Korea, e-mail: \{xuanloc088, 
salman0335, glideslope, cshong\}@khu.ac.kr. }
\thanks{Yan Kyaw Tun is with the Department of Electronic Systems, Aalborg University, 2450 København SV, Denmark, e-mail:{\{ykt\}@es.aau.dk}.
}
\thanks{Zhu Han is with the Department of Electrical and Computer Engineering at the University of Houston, Houston, TX 77004 USA, and also with the Department of Computer Science and Engineering, Kyung Hee University, Seoul, South Korea, 446-701, e-mail:  hanzhu22@gmail.com.}
\vspace{-0.3in}}

%
%

\markboth{Journal of \LaTeX\ Class Files,~Vol.~14, No.~8, August~2015}%
{Shell \MakeLowercase{\textit{et al.}}: Bare Demo of IEEEtran.cls for IEEE Journals}
%



\maketitle

\begin{abstract}

The sixth-generation (6G) non-terrestrial networks (NTNs) are crucial for real-time monitoring in critical applications like disaster relief. However, limited bandwidth, latency, rain attenuation, long propagation delays, and co-channel interference pose challenges to efficient satellite communication. Therefore, semantic communication (SC) has emerged as a promising solution to improve transmission efficiency and address these issues. In this paper, we explore the potential of SC as a bandwidth-efficient, latency-minimizing strategy specifically suited to 6G satellite communications. While existing SC methods have demonstrated efficacy in direct satellite-terrestrial transmissions, they encounter limitations in satellite networks due to distortion accumulation across gateway hop-relays. Additionally, certain ground users (GUs) experience poor signal-to-noise ratios (SNR), making direct satellite communication challenging. To address these issues, we propose a novel framework that optimizes gateway hop-relay selection for GUs with low SNR and integrates gateway-based denoising mechanisms to ensure high-quality-of-service (QoS) in satellite-based SC networks. This approach directly mitigates distortion, leading to significant improvements in satellite service performance by delivering customized services tailored to the unique signal conditions of each GU. Our findings represent a critical advancement in reliable and efficient data transmission from the Earth observation satellites, thereby enabling fast and effective responses to urgent events. Simulation results demonstrate that our proposed strategy significantly enhances overall network performance, outperforming conventional methods by offering tailored communication services based on specific GU conditions.
\end{abstract}
\IEEEpeerreviewmaketitle
\section{Introduction}
\IEEEPARstart{T}{he} non-terrestrial networks (NTNs), specifically satellite access networks (SANs), have emerged as a vital solution for achieving large-scale connectivity and seamless coverage, which are critical for next-generation wireless networks \cite{NTN+AI-motivation}. Unlike terrestrial communication systems, NTNs face challenges such as high launch costs, dependence on line-of-sight (LoS) transmission, and long propagation distances. To address these challenges, low Earth orbit (LEO) satellite\footnote{Hereinafter, LEO satellite will be considered as satellite unless otherwise stated.} systems like Iridium, Globalstar, OneWeb, Starlink, and Telesat have been deployed \cite{various_LEO_satellites}. These systems initially used transparent satellites that amplified signals but now rely increasingly on regenerative satellites \cite{regenerative_satellites}. Regenerative satellites function as base stations with onboard processing capabilities, such as modulation and coding, and support inter-satellite links within LEO constellations—an essential feature for global networking.

The high mobility of satellites leads to performance degradation, prompting research into improving the reliability and adaptability of NTNs \cite{enhancing-transmission-reliability-adaptability, ntn-fading}. Techniques like robust beamforming, adaptive modulation, and interference mitigation are being explored to enhance transmission performance. Large satellite constellations, combined with the diversity of services and wide coverage areas, introduce further complexity. Additionally, aggressive frequency reuse in satellite systems, aimed at increasing spectrum efficiency, exacerbates co-channel interference (CCI). Addressing the unpredictable nature of CCI, caused by satellite movement, remains an ongoing challenge. While hybrid automatic repeat request (HARQ) mechanisms are sometimes employed to manage transmission errors, these are limited by the substantial and variable delays between satellites and ground stations \cite{HARQ}. As a result, deep learning has been proposed as a solution to tackle these complex challenges in NTNs \cite{DL_SAN}. 

Semantic communication (SC) has shown significant potential in reducing bandwidth consumption and improving performance in challenging channel conditions \cite{SC-MOT}. By focusing on transmitting the meaning (semantics) of data rather than raw data, SC aligns with the vision of \emph{Intelligence as a Service} in 6G networks. Recent advances have demonstrated the utility of SC in addressing NTN-specific challenges, such as satellite computation offloading, wireless UAV control, and remote spectrum sensing. However, further research is required to enhance the integration of SC into NTNs \cite{sat_AI}. The SC prioritizes task-oriented performance over traditional bit-level accuracy and has proven effective in addressing connectivity and low-latency issues in 6G. This approach is made possible by advancements in deep learning models, which enable efficient embedding of semantic information into the communication process \cite{10554663}. The integration of SC into NTNs presents an innovative path forward, though significant challenges remain in optimizing satellite transmission using this paradigm.

This paper explores the potential of integrating SC with SANs, a powerful approach that focuses on conveying the meaning (semantics) of information rather than just the data itself \cite{SC-NTN-ref1}. This integration offers several advantages i.e., SC has the potential to reduce communication overhead. Additionally, bridging the digital divide can ensure equitable access to high-quality communication services in underserved regions through integration with terrestrial networks. These advancements pave the way for a global \textit{Intelligence as a Service} network, fostering a more connected and intelligent world. However, traditional hop-relaying techniques in satellite communication face limitations like high power consumption and error accumulation over long distances. Therefore, we introduce a novel architecture for NTNs based on SC\footnote{In this research article, NTNs refer specifically to SANs.}, showcasing its promising applications. Integrating SC within NTN is not without its hurdles, i.e., we identify primary challenges, e.g., including hop-based SC across multiple satellite constellations and effective denoising of data at the receiver end. This research aims to address these challenges, paving the way for more efficient and reliable NTN communications. 

To address the above-mentioned challenges, we investigate the application of SC in satellite-based hop relay networks. Our approach considers both direct access for ground users (GUs) with favorable signal conditions and selective hop relaying for enhanced reliability and fidelity. While most existing SC research focuses on single-hop terrestrial communication, this paper ventures into a novel concept where hop-relaying satellite networks with denoising are implemented. We acknowledge valuable contributions to SC in terrestrial networks, such as physical-layer network coding with semantic decoders \cite{two-waychannel}, cooperative transmission with half-duplex relays \cite{bian2022deep}, and autoencoder-based SC systems with semantic forward protocols \cite{relaychannel_AE}. This research seeks to bridge the gap by investigating SC in satellite hop relay networks. Despite these advancements, SC in practical cooperative 3D networks or remote wireless sensor networks remains under-explored, especially where the destination is far from the source. This necessitates satellite-based relaying to ensure remote coverage and avoid deep fading through line-of-sight (LoS) connections.

Motivated by this gap, this paper investigates SC on hop relay networks and evaluates the impact of hop forwarding on the fidelity of semantic information. To the best of our knowledge, this is the first evaluation of semantic attenuation loss of semantic information on a satellite hop relay network. The major contributions of this work are summarized as follows:
\begin{itemize}
    \item In this paper, we consider an SC system tailored for satellite architectures that serve multiple GUs. Recognizing that the high altitude of satellites can impact the SNR for GUs, we propose a gateway-assisted approach to enhance communication quality for these GUs.
    \item We leverage the robust capacity and extensive resources of the gateway in NTN communications to mitigate environmental noise and meet the high QoS demands of GUs. Specifically, the received power at the receiver influences the signal-to-noise ratio (SNR), with the gateway delivering a higher quality signal compared to mobile GU devices.
    \item Next, we formulate an optimization problem to minimize the total communication overhead for all GUs while ensuring their reconstruction quality requirements are met. This is accomplished by optimizing the pathway variables and carrier frequencies. To address this problem efficiently, we introduce a two-stage discrete whale optimization algorithm (D-WOA), which delivers a sub-optimal solution within a polynomial time frame, crucial for adapting to the satellite's dynamic environment.
    \item Finally, we evaluate the system's performance with a standard satellite dataset, comparing it against various optimization benchmarks. Additionally, we assess the effectiveness of the gateway-assisted hop approach in producing high-quality images in a high-noise environment.
\end{itemize}

The remainder of this paper is organized as follows. Section \ref{rel_work} summarizes the related work, while Section \ref{sys_model} provides the system model. Section \ref{prob_form} details the problem formulation, and Section \ref{sol_approach} introduces the proposed solution, including the network structure and training process. Section \ref{sim_results} showcases simulations performed on the satellite-based dataset. Finally, Section \ref{conc} concludes the paper.
\section{Related Works}
\label{rel_work}
\subsection{Non-Terrestrial Networks}
Emerging terrestrial communication infrastructure supports advanced services but is costly, limiting deployment to urban areas \cite{NTN-1}. To achieve seamless 6G connectivity, NTNs are gaining attention for their flexible, high-throughput, and cost-effective deployment, especially in remote or harsh environments. The development of satellite constellations and UAV base stations (BSs) promises high-quality network connections and diverse services for remote GUs \cite{NTN-2, NTN-3}. The growing demand for various applications in intelligent transportation systems (ITS) underscores the limitations of conventional terrestrial networks for remote nodes such as airplanes and ships. The study in \cite{NTN-4} introduces a novel approach that utilizes SANs along with LEO and cube satellites as multi-access edge computing (MEC) servers for data offloading and computation services. It focuses on optimizing offloading task selection and resource allocation to tackle the challenges of joint delay and rental price minimization.
\vspace{-0.1in}
\subsection{Semantic Communication}
As 5G technology becomes widespread, attention shifts to the next generation of communication systems, where current methods are nearing Shannon's capacity limit despite advancements in encoding, decoding, and modulation. However artificial intelligence (AI) plays a crucial role in evolving these technologies, particularly in the emerging paradigm of SC, which focuses on \textit{semantic-meaning passing}. SC extracts and transmits ``meanings" using a shared knowledge base (KB) for accurate interpretation at the receiver \cite{SC-1}. The innovations i.e., semantic similarity check (SSC) and semantic combining (SeC) methods have been introduced to enhance semantic information recovery and merge semantic data from multiple paths. In contrast, a new metric, semantic energy efficiency (SEE), balances information recovery with energy consumption \cite{SC-2}. Further advancements include a reinforcement learning-based adaptive semantic coding (RL-ASC) approach for bandwidth-sensitive, semantic-rich image data, which optimizes rate-semantic-perceptual performance and employs a GAN-based decoder to reconstruct images \cite{SC-3}. Another study enhances SC by emphasizing the importance of semantic information transmission and recovery, introducing the concept of ``flow of intelligence" and semantic slice models (SeSM) for flexible adaptation under varying conditions. This work also proposes a layer-based SC system for images (LSCI) and a novel evaluation metric, semantic service quality (SS) \cite{SC-4}. A distributed training framework for semantic communication is developed in \cite{10531097} without violating the privacy of the training participants. In \cite{10388241}, a semantic forward relaying framework is developed to reduce the payload of the relay-destination link.
\vspace{-0.1in}
\subsection{Semantic Communication Based Non-Terrestrial Networks}
The authors in \cite{SC-NTN-1} propose a novel SC-assisted satellite-borne edge cloud (SemCom-SEC) framework for computation offloading of resource-limited GUs. This includes an adaptive pruning-split federated learning (PSFed) method to update the semantic coder, ensuring training convergence, accuracy, privacy, and reduced training delay and energy consumption. For computational tasks, the goal is to minimize delay and energy consumption while maintaining privacy and fairness. Additionally, accurate spectrum sensing is crucial for spectrum coexistence between terrestrial and NTNs, with UAVs detecting spectrum availability in remote areas. To address UAV computing limitations and high latency, a semantic-oriented framework for real-time spectrum sensing is proposed \cite{SC-NTN-2}, enabling UAV-BS collaborative detection to reduce data size and using deep reinforcement learning (DRL) for optimal UAV trajectory tracking of primary GUs (PUs). The paper in \cite{SC-NTN-3} addresses the challenge of optimizing energy efficiency in space-air-ground integrated networks (SAGINs) using probabilistic SC (PSC). The study formulates an optimization problem to minimize energy consumption across communication and computation tasks in a satellite-UAV-ground terminal system. 

Similarly, the study in \cite{SC-NTN-4} proposes a foundation model (FM)-based SC framework for satellite systems, termed FMSAT, to enhance transmission efficiency in challenging environments. The framework utilizes FM-based segmentation and reconstruction to reduce bandwidth requirements and accurately recover semantic features despite high noise and interference. An adaptive encoder-decoder is introduced to safeguard critical features and minimize re-transmissions, while a novel error detection method at both the satellite and gateway addresses long propagation delays. In our recent study \cite{our_globecom}, we present an SC-based approach aimed at optimizing latency in the satellite imagery downlink process for Earth observation, focusing solely on direct communication scenarios. By harnessing the inherent efficiency of SC, the proposed solution significantly reduces communication overhead and enhances spectrum utilization. This approach demonstrates particular relevance for 6G NTNs, offering a promising strategy to address the unique challenges of latency and resource management in these advanced communication systems.
\vspace{-0.2in}
\subsection{Challenges in the Literature and Rationale for Proposal}
Previous studies have underscored the increasing significance of NTN in mobile communication, particularly highlighting their extensive coverage and seamless connectivity \cite{SC-NTN-Chal-1, SC-NTN-Chal-2}. These works propose a SC-based NTN architecture and outline its potential applications. However, they fail to address several critical challenges, including low-complexity feature extraction for diverse GUs, the implementation of a denoising mechanism at the reception gateway, and dynamic network resource allocation based on GU channel conditions. In response, our proposed work investigates the application of AI-driven low-complexity swin-transformers to effectively extract meaningful features from raw data. Additionally, we explore intelligent resource management strategies tailored to individual GU conditions to mitigate these challenges, thereby enabling lightweight and efficient SC in NTN. Furthermore, we implement a denoising mechanism utilizing the swin-transformer architecture to eliminate noise from coded information before transmission. This approach is essential for GUs who may struggle to process the raw data, which could otherwise lead to significantly degraded QoS in the network.
\begin{figure*}[t]
    \centering
    \includegraphics[width=0.75\textwidth, height=4in]{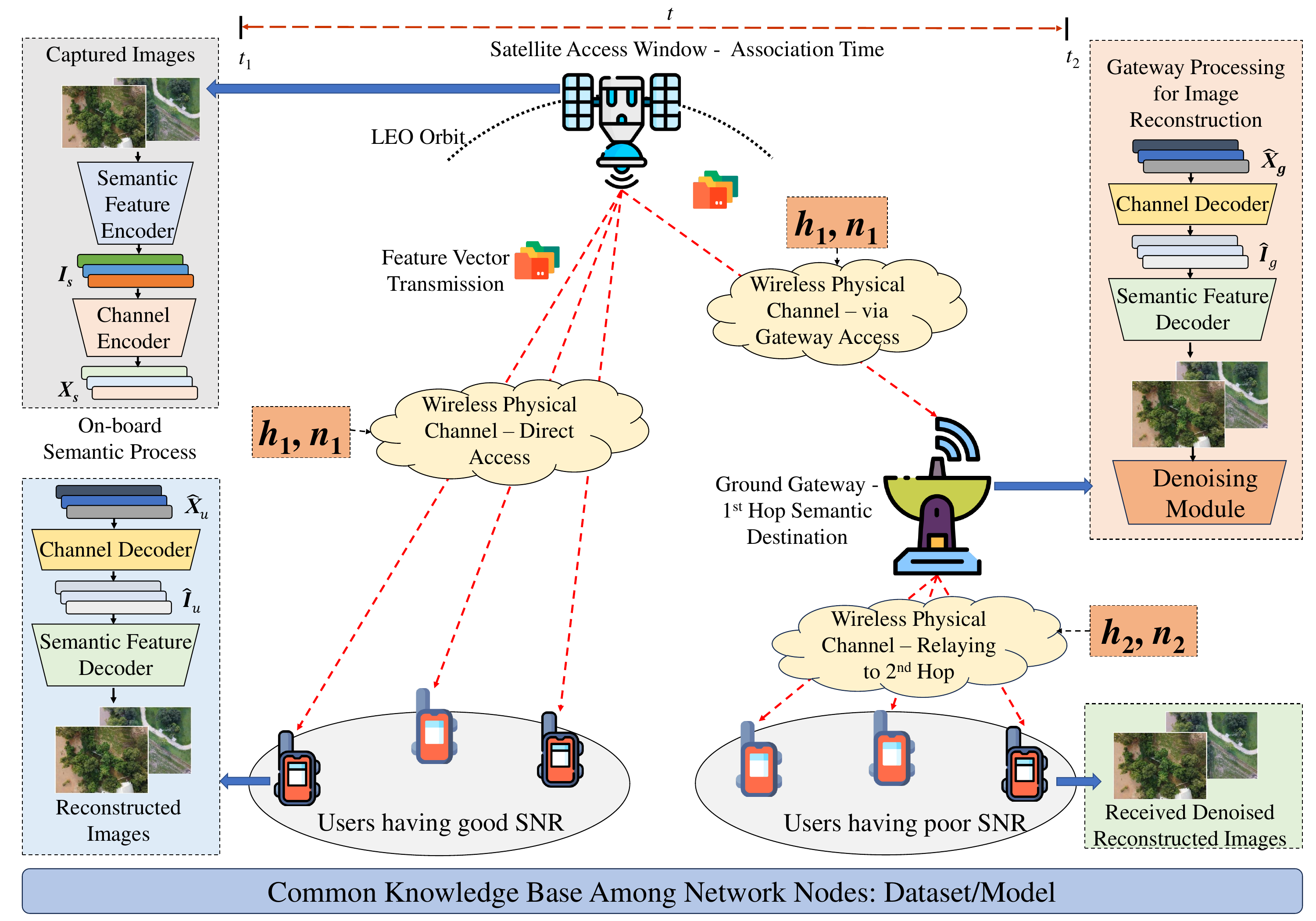}
    \caption{The architectural framework for facilitating satellite-based semantic communication to the diverse remote GUs.}
    \label{system_model_fig}
    \vspace{-0.2in}
\end{figure*}
\section{System Model}
\label{sys_model}
\subsection{Network Model}
As illustrated in Fig.~1, the proposed SC system comprises satellite $s$ gateway $g$, and a set $\mathcal{U}$ of $U$ GUs, i.e., $u, u' \in \mathcal{U}$, where $u$-indexed GUs are those whose signal-to-noise ratio (SNR) favorable for direct satellite-based SC and $u'$-indexed GUs are those which cannot directly associate with satellite for SC due to their poor SNR, thus $u \cup u' = U$. Let's define another joint set $\mathcal{M}$ of $M$ composed of all good SNR GUs and a gateway known as direct ground nodes. Here, we consider the captured image reconstruction task as the service of the system, where the satellite encodes the semantic features within the image and transmits them to GUs within the network domain. Then, the GUs try to reconstruct the original image from the received signal, and each GU will have different requirements for the QoS, such as time or image quality. Due to the mobility of the satellite and the dispersed locations of the GUs, the connections between the satellite and GUs are dynamic, impacting the channel quality. When the channel conditions between the satellite and a GU are favorable, the GU can achieve high-quality image reconstruction. Conversely, at elevated noise levels, the decoders at the GUs struggle to reconstruct images with high fidelity. To overcome this challenge, we propose utilizing a ground gateway as an intermediary hop and denoising node in the SC process. Specifically, the satellite transmits the image to the gateway, which benefits from virtually unlimited computational power and energy resources. At the gateway, a DL-based denoising model is employed to mitigate the noise effects before relaying the refined semantic signal to the intended GUs. Additionally, when the gateway itself serves as the target GU, it can fully reconstruct the image with high accuracy.\footnote{It is assumed that since the gateway possesses substantial computing power, it is more efficient to reconstruct the image/data at the gateway rather than relying on low SNR GUs. This approach ensures optimal utilization of network resources and enhances service quality.} 
\vspace{-0.1in}
\subsection{Satellites-based Semantic Feature Extraction Model}
Similar to standard SC systems, the transmitter at the satellite comprises a semantic encoder and a channel encoder, which will be jointly optimized. The semantic encoder is responsible for extracting the important features within the image $I$, which significantly reduces the transmitted data and saves more bandwidth resources. We denote the size of the original image as $n$ and $I \in $ $\mathbb{R}^n$, the value of $n$ is calculated by the multiplication among its height, width, and the number of color channels. The semantic features of the image $I$ can be denoted as:
\begin{equation}
    F_{I}= E_{\alpha}(I),~\forall I \in \mathbb{R},  \label{feature}
\end{equation}
where $E_{\alpha}(\cdot)$ denotes the semantic encoder and $\alpha$ is its parameter. The semantic features are then compressed by the channel encoder, which lowers the transmission length and protects them from the physical noise from the environment. To effectively combat the noise, the channel encoder will take the $\Gamma$ SNR value as a condition in the process and output the encoded semantic symbol as: 
\begin{equation}\label{transmitsignal}
    X_{I}= C_{\beta}(F_{I}|\Gamma), ~\forall \Gamma \in \mathbb{SNR},
\end{equation}
where $C_{\beta}(\cdot)$ indicate the channel encoder and its corresponding network parameters $\beta$. The size of the encoded semantic symbol $X_{I}$ is expressed as $b$ and the ratio between the size of the encoded semantic symbol and the original image size is calculated by $b/n$. 
\vspace{-0.1in}
\subsection{Satellites-based Semantic Feature Transmission Model}
We consider an orthogonal frequency-division multiple access (OFDMA) due to its flexibility in allocating limited network resources. This allows satellites to be assigned subcarriers based on their specific data requirements, efficiently handling dynamic GU channel conditions. Additionally, OFDMA inherently avoids inter-GU interference within the spectrum. A set $\mathcal{K}$ of $K$ subcarriers is considered to allocate each GU if its SNR is above the QoS threshold (i.e., good SNR). Otherwise, the gateway will receive this semantic signal and will process and relay it to the end GUs after the reconstruction and denoising process. Let $\gamma^{k}_{m}$ denote the subcarrier assignment variable for each GU $u$ or gateway $g$ is:
\begin{equation}
    \gamma_{s,m}^k \in \{0,1\}, \label{association_constraint_1_1}
\end{equation}
i.e., 
\begin{equation}
    \gamma_{s,m}^k = 
        \begin{cases}
         1,     & \textrm{if}~\textrm{subcarrier}~k~\textrm{utilized}~\textrm{by}~\textrm{any}~\textrm{direct}\\
                &\textrm{connection,}~\textrm{i.e.,}~\textrm{GU~$u$}~\textrm{or}~\textrm{gateway}~\textrm{$g$,}\\
         0,     & \textrm{otherwise}.
        \end{cases}
\end{equation}

We also provide the condition that only one subcarrier $k$ is assigned to GU $u$ or gateway $g$ at each time slot to avoid network interference:
\begin{equation}
   \sum_{m=1}^{M} \gamma_{s,m}^k \leq 1. \label{association_constraint_1_2}
\end{equation}
Similarly, each node $m$ can utilize at most a single subcarrier, which is given as follows:
\begin{equation}
    \sum_{k=1}^{K} \gamma_{s,m}^k \leq 1. \label{association_constraint_1_3}
\end{equation}  

Thus, the downlink rate from the satellite to the GU $u$ or gateway $g$ on subcarrier $k$ to transmit semantic feature information $F_{I,k}$ as:
\begin{equation}
  R_{s,m}^k =   B_{s,m}^k \log_2 \big( 1 + \Gamma_{s,m}^k\big), ~ ~\forall m \in \mathcal{M}, ~k \in \mathcal{K},
   \label{rate_1}
\end{equation}
where $B_{s,j}^k$ is the subcarrier bandwidth. Moreover, the SNR $\Gamma$ between the satellite and the ground node $m$ can be defined as follows:
\begin{equation}\label{SNRvalue}
    \Gamma_{s,m}^k = G_s G_m P_s \Big(\frac{c}{4\pi d_{s,m} f_{s,m}^k N_0} \Big)^2,
\end{equation}
where $G_s$ and $G_m$ denote the wireless power antenna gains for the satellite and GU/gateway, respectively. $P_s$ is the satellite transmit power, $c$ indicates the speed of light, $f_{s,m}^k$ denotes the carrier frequency, $d_{s,m}$ denotes the distance between the satellite $s$ and the GU $u$ or gateway $g$, and $N_0$ denotes the spectral density of the noise power to capture the interference generated by the satellites in other constellations which utilize the same bandwidth spectrum and subcarrier $k$. 
\vspace{-0.2in}
\subsection{GU Semantic Feature Reception and Reconstruction}
The GUs can receive the semantic encoded feature from the satellite in two pathways: direct connection and indirect connection via the gateway. This pathway will be determined by environmental noise and the GU demand for image quality. In both cases, the GU decodes the encoded semantic symbol using its semantic and channel decoders. 
\subsubsection{Direct GU connections}
With the encoded semantic symbol transmitted through a wireless channel, environment noise is inevitable, and the received signal at the GU is:
\begin{equation}\label{noiseSU}
    Y_{I,u}= X_{I}+ N_{u}
\end{equation}
where $N_{u}$ denotes the channel noise, whose element follows the Gaussian distribution $\mathcal{N}(0,\sigma^{2}\boldsymbol{I})$. Here we consider both the additive white Gaussian noise (AWGN) channel and the Rayleigh fading channel.
\begin{equation}
    Y_{I,u}= H_{I,u} X_{I} + N_{u},
\end{equation}
where $H_{I,u}$ denotes the fading coefficient of the Rayleigh channel between communication devices. Initially, GU $u$ decodes the received signal with the channel decoder:
\begin{equation}
    \hat{F}_{I,u}=C^{-1}_{\phi_u}(Y_{I,u}),
\end{equation}
where $C^{-1}_{\phi_u}(\cdot)$ represent the channel decoder at GU $u$ and the corresponding parameters $\phi_{u}$. Then the semantic decoder takes the semantic feature of the image $I$ as input and reconstructs the original image as:
\begin{equation}
    \hat{I}_{u}= D_{\theta_{u}}(\hat{F}_{I,u}),
\end{equation}
where $D_{\theta_{u}}(\cdot)$ denote the semantic decoder of GU $u$ and its parameter $\theta_{u}$.  
\subsubsection{Indirect GU connections via Gateway}
\label{indirect_comm}
In extremely harsh environmental conditions, GUs cannot achieve high-quality image reconstruction due to their models and power limitations. To address this, the satellite actively transmits signals to a gateway, relaying these signals to the GU. As shown in the equation (\ref{SNRvalue}), with the same environment noise and approximate distance, the gateway is equipped with powerful antenna gains, leading to better-received signals than the GU. Additionally, leveraging the gateway's abundant resources, we propose a DL-based denoising model to mitigate noise before relaying the signal to the GU $u'$. Details of the denoising model are provided in \ref{denoiseandrelay}. Here, we focus on the link between the gateway $g$ and the GU $u'$. Lets $\chi^j_{g,u'}$ denotes the subcarrier assignment variable for each GU $u'$ is:
\begin{equation}
    \chi^j_{g,u'} \in \{0,1\}, \label{association_constraint_2_1}
\end{equation}
i.e., 
\begin{equation}
    \chi^j_{g,u'} = 
        \begin{cases}
         1,     & \textrm{if}~\textrm{subcarrier}~j~\textrm{utilized}~\textrm{by}~\textrm{any}~u~\textrm{indirect}~\textrm{GU,}\\
         0,     & \textrm{otherwise}.
        \end{cases}
\end{equation}

We also provide the condition that only one subcarrier $j$ is assigned to each GU at each time slot to avoid network interference:
\begin{equation}
   \sum_{u'=1}^{U-u} \chi^j_{u'} \leq 1. \label{association_constraint_2_2}
\end{equation}
Similarly, each GU $u'$ can utilize at most a single subcarrier, which is given as follows:
\begin{equation}
    \sum_{j=1}^{J} \chi^j_{u'} \leq 1. \label{association_constraint_2_3}
\end{equation}   The downlink rate from the gateway $g$ to the GU $u'$ on subcarrier $j$ to transmit the data of reconstructed image $I$ as:
\begin{equation}
   R_{g,u'}^j =  B^{u'}_g \log_2 \big( 1 + \Gamma_j \big),  ~~\forall u' \in \mathcal{U}-u, ~j \in \mathcal{J}, 
   \label{rate_2}
\end{equation}
where $B_g$ is the subcarrier bandwidth. Moreover, the SNR $\Gamma$ between the gateway and the GU can be defined as follows:
\begin{equation}
    \Gamma_{g,u'}^j = G_g G_u P_g \Big(\frac{c}{4\pi d_{g,u'} f_{g,u'} N_1} \Big)^2,
\end{equation}
where $G_g$ and $G_{u'}$ denote the wireless power antenna gains for the gateway and GU, respectively. $P_j$ is the gateway transmit power, $c$ indicates the speed of light, $f_c$ denotes the carrier frequency, $d_{g,u'}$ denotes the distance between the gateway and the GU.
\vspace{-0.2in}
\subsection{Integrated Hop Gateway for Processing and Relaying}\label{denoiseandrelay}
The assistance provided by the gateway to the GU is twofold: leveraging a powerful antenna and denoising the signal using unlimited computing resources. Equation (\ref{SNRvalue}) shows that the SNR is influenced not only by environmental noise but also by the transmitter and receiver power and the distance between them. GU constrained by limited power, typically experiences low SNR in harsh physical environments and over long distances. In contrast, when the gateway receives the signal under the same noise conditions ($N_{0}$) and a similar distance ($d_{s,u}$), the gateway’s powerful antenna significantly enhances the SNR. Consequently, the gateway can capture a signal that is closer to the original transmitted signal compared to the GU device. Furthermore, utilizing its vast computing resources, the gateway can employ a deep learning-based denoising module to further reduce environmental noise before relaying the signal to the GU. Besides, the gateway can be a targeted receiver, which can generate high-quality images due to its powerful model and antenna gain.\\

\subsubsection{Signal Relay Mode for the GUs}
Under this mode, the gateway first decodes the received signal by employing its joint source-channel decoder. 

\begin{equation}
    \hat{I}_{g}=D_{\theta_{g}}(C^{-1}_{\phi_g}(Y_{I,g}|\Gamma)),
\end{equation}
where $Y_{i,g}$ denotes the received signal and $\hat{I}_{g}$ is the reconstructed image at the gateway. This image is reconstructed here, which can be beneficial not only to the high powers of the antenna here but also to the computing resource. To be specific, a denoising model refines the reconstructed image before putting it through the encoding process and transmit to the receiver. The denoising model requires a high computing capacity to implement, which is infeasible for mobile GUs. This process can be denoted as the following equation:
\begin{equation}
    \hat{I}_{\mathrm{den}}= DEN_{\eta}(\hat{I}_{g}),
\end{equation}
where $DEN_{\eta}(\cdot)$ denote the denoising model with the parameter $\eta$. 
After finishing the denoising process, the gateway transmits the denoised signal to the GU, which also counters the noise from the environment. However, the signal will be less severely affected due to the relatively small distance compared to the satellite-GU distance. The encoding process at the gateway is represented as:
\begin{equation}
    X_{I,g,u}= C_{\beta_{g}}(E_{\alpha_g}(\hat{I}_{\mathrm{den}})|\Gamma),
\end{equation}
where $X_{I,g,u}$ is the encoded signal at the gateway. GU $u$ received the signal $Y_{I,g,u}$, which also experiences the environment noise and follows the formula as equation (\ref{noiseSU}). Then, GU $u$ decodes the signal to acquire the image. 
\subsubsection{Semantic Image Reconstruction}  
If the gateway is the intended recipient, it can easily reconstruct the image using the joint source-channel decoder. Additionally, it's worth noting that the denoising process is optional, which depends on the GU's and gateway's quality requirements.
\vspace{-0.2in}
\section{Latency Minimization Problem Formulation}
\label{prob_form}
Based on the encoded semantic features obtained in (\ref{transmitsignal}), the achievable data rates calculated in (\ref{rate_1}), we can calculate the latency of directly transmitting the features from satellite $s$ in subcarrier $k$ to the GUs or gateway $m$ is:
\begin{equation}
  t_{s,m}^k = \frac{l_{X}}{R_{s,m}^k}, ~\forall m \in \mathcal{M},
\end{equation}
where $l_{X}$ denotes the length of the transmitted data. For GUs who receive the signal through the gateway indirect paths, the latency is the sum of two transmission processes: \textit{latency from satellite to gateway}, \textit{latency from gateway to GU}. The latency is calculated as:
\begin{equation}
t_{s,u'}= \frac{l_{X}}{R_{s,m}^k} + \frac{l_{X}}{R_{g,u'}^j}, \forall u' \in \mathcal{U}-u, ~m=g,
\end{equation}
where $R_{s,m}^k$ is only that satellite $s$ is connected with gateway $m$ via subcarrier $k$ for relaying towards bad SNR GUs $u'$. 

Leveraging the aforementioned system model, we aim to minimize the average latency for transmitting semantic features from satellite $s$ to GUs based on the optimization of expected received image quality $\mathop{\mathbb{E}}(\mathrm{PSNR}_k(\boldsymbol{X}))$. However, this optimization must adhere to the network model's practical constraints. Specifically, sufficient semantic information must be transmitted within the access window to allow the direct access GUs or gateway to reconstruct the original send image. Additionally, reliability and SC-QoS must be ensured by guaranteeing a minimum PSNR threshold that meets the desired image recovery quality. Under these assumptions, the average latency from  satellite to gateway can be given as:
\begin{equation}
 \Psi_1 (l_{X}, \boldsymbol{\gamma}) = \frac{1}{M}  \sum_{m \in \mathcal{M}} \sum_{k\in \mathcal{K}} \gamma_{s,m}^k t_{s,m}^k.
\end{equation}
Similarly, the average latency from the gateway to GU can be given as:
\begin{equation}
  \Psi_2(l_{X}, \boldsymbol{\chi}) = \frac{1}{U-u} \sum_{u' \in \mathcal{U}-u} \sum_{j \in \mathcal{J}} \chi^j_{g,u'}t_{g,u'}. 
\end{equation}
Then the total latency could be: 
\begin{equation}
    \boldsymbol{O}(\boldsymbol{S},\boldsymbol{G},\boldsymbol{U}, \boldsymbol{\gamma}, \boldsymbol{\chi}) = \Psi_1 + \Psi_2.
\end{equation}
The latency minimization problem under the recovered image quality optimization and loss function minimization can be formulated as:
\begin{mini!}|s|[2]<b>                
		{\substack{\boldsymbol{S},\boldsymbol{G},\boldsymbol{U}, \boldsymbol{\gamma}, \boldsymbol{\chi} }  } 
		{ \boldsymbol{O}(\boldsymbol{S},\boldsymbol{G},\boldsymbol{U}, \boldsymbol{\gamma}, \boldsymbol{\chi})  }{\label{P1}}{\textbf{P1:}}	
		\addConstraint{\gamma^{k}_{m} \in \{0,1\}, \quad \forall k \in \mathcal{K}, ~\forall m \in \mathcal{M}} {\label{prob_association_constraint_1} }
		\addConstraint{\sum_{m=1}^{M} \gamma^{k}_{m} \leq 1, \quad \forall k \in \mathcal{K}} {\label{prob_association_constraint_2}}
		\addConstraint{ \sum_{k=1}^{K} \gamma^{k}_{m} \leq 1, \quad \forall m \in \mathcal{M}} {\label{prob_association_constraint_3}}
        \addConstraint{\chi^j_{u'} \in \{0,1\}, \quad \forall j \in \mathcal{J}, ~\forall u' \in \mathcal{U}-u} {\label{prob_association_constraint_4} }
		\addConstraint{\sum_{u'=1}^{U-u} \chi^j_{u'} \leq 1, \quad \forall j \in \mathcal{J}} {\label{prob_association_constraint_5}}
		\addConstraint{ \sum_{j=1}^{J} \chi^j_{u'} \leq 1, \quad \forall u' \in \mathcal{U}-u} {\label{prob_association_constraint_6}}
		\addConstraint{\mathbb{E}(\mathrm{PSNR}_u(\boldsymbol{X}))} \geq \Psi_u, \quad \forall u \in \mathcal{U}-u' {\label{PNSR_Constraint_7}}
  		\addConstraint{\mathbb{E}(\mathrm{PSNR}_{u'}(\boldsymbol{X}))} \geq \Psi_{u'}, \quad \forall u' \in \mathcal{U}-u {\label{PNSR_Constraint_8}}
  	\addConstraint{0 \leq \gamma^{k} t_{s,m} \leq T_m, \quad       \forall m \in \mathcal{M}},      {\label{Access_window_Constraint_8}}
\end{mini!}
where $\boldsymbol{S=(\alpha,\beta})$ comprise the parameters of the semantic and channel encoders at the transmitter, while $\boldsymbol{G=(\theta_g,\phi_g,\eta,\alpha_g,\beta_g)}$ and $\boldsymbol{U=(\theta_u,\phi_u)}$ denotes the network parameters of the gateway and GU $u$, respectively. Constraints (\ref{prob_association_constraint_1}), (\ref{prob_association_constraint_2}), and (\ref{prob_association_constraint_3}) guarantee that the satellite $s$ will assign only a single subcarrier $k$ to high-end ground nodes $j$ (i.e., direct GU $u$ or gateway $g$) for feature transmission. Every subcarrier could only be utilized by a single ground node $m$ at each time slot for semantic features reception.
Constraints (\ref{prob_association_constraint_4}), (\ref{prob_association_constraint_5}), and (\ref{prob_association_constraint_6}) are second-stage constraints that guarantee that the gateway $g$ will assign only a single subcarrier $j$ to low-end ground nodes $u'$ for denoised feature transmission, and every subcarrier could only be utilized by a single ground node $u'$ at each time slot for denoised semantic features reception. Both constraints (\ref{PNSR_Constraint_7}) and (\ref{PNSR_Constraint_8}) ensure that the semantic features should be reliable and meet the SC-QoS threshold for every GU in a set $\mathcal{U}$. Constraint (\ref{Access_window_Constraint_8}) ensures the communication efficiency of satellites within the access window, i.e., the coverage period of the satellite $s$ to ground nodes $m$.
It can be observed that the formulated problem is MINLP, which is difficult to solve due to its being NP-hard. To address this problem, we provide a solution approach in the next section.
\vspace{-0.1in}
\section{Proposed Algorithm}
\label{sol_approach}
As shown in the formulation, the proposed problem depends on the deep learning parameters of the encoders, decoder, and denoise modules, which is similar to most other SC studies. Thus, optimizing the parameters of these modules is the top priority so that the SC system can effectively extract the image feature and efficiently transmit it over a wireless environment. Then, based on the optimized parameters of all the modules and the GU requirements, the transmit pathway will be decided to meet the demands of all GUs within the lowest communication time.
\vspace{-0.1in}
\subsection{Parameters Optimization of Semantic Communication}
In this subsection, we introduce the architecture of each encoder/decoder for the channel or source coding process, the denoising model, and finally, the optimization algorithm. To make it easier for the reader to follow, we will refer to the encoder/decoder pair as the coder from now on.
\subsubsection{Source Coder}
\begin{figure}[t]
    \centering
    \includegraphics[width=0.8\columnwidth]{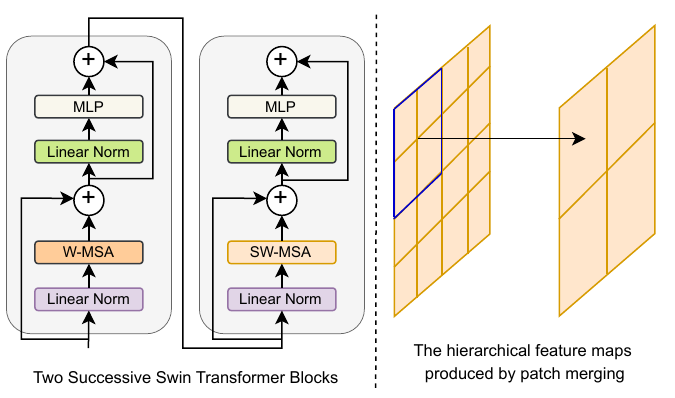}
    \caption{Two Swin transformer blocks and hierarchical feature maps.}
    \label{Swin}
   \vspace{-0.3in}
\end{figure}

Semantic extraction from images is a challenging task and thus requires a complex model to capture them. Therefore, we adopt the Swin Transformer model as the backbone architecture of the semantic coder for the transmitter and all the network receivers. The model is novel due to its excellent performance capacity in extracting the latent feature within the data while requiring a competitively low computation complexity compared to other transformers or stacking conventional convolution techniques. It first divides an image into non-overlapping small patches and applies the transformer mechanism on each patch. The local self-attention within each image patch costs significantly lower computing operations compared with the global attention proposed in the Vision Transformer. In addition, to create the connection among patches, every normal window is followed by a shifted window, which is implemented to capture the dependency correlation of the image, thus extracting the semantic feature effectively. Finally, a hierarchical map is constructed by the patch merging layer; specifically, the model concatenates each group's feature maps of $2 \times 2 $ neighboring patches from the previous stage and then puts the result in the next Swin Transformer blocks as shown in Fig.~\ref{Swin}. The formal formulation for two consecutive transformer blocks is presented as follows:
\begin{align*}
    &\tilde{k}^{l}= \windowhyphen(LN(k^{l-1})+k^{l-1},\\
    &k^{l}= \mathrm{MLP}(\mathrm{LN}(\tilde{k}^{l}))+\tilde{k}^{l},\\
    &\tilde{k}^{l+1}= \shiftwindowhyphen(\mathrm{LN}(k^{l})+k^{l},\\
    &k^{l+1}= \mathrm{MLP}(\mathrm{LN}(\tilde{k}^{l+1}))+\tilde{k}^{l+1}. 
\end{align*}
For the transmitter, the output of the final block will be encoded by the channel encoder. At the receiver site, the semantic decoder will take the output of the channel decoder as input and reconstruct the image. Thus, instead of merging patches like in the encoder, the architecture of the semantic decoder will be inversed of the encoder, such as patch splitting instead of patch merging. We consider the difference in computing capacity between the GU and BS by stacking more transformer blocks for the BS while keeping the minimum Swin transformer blocks for all GU devices. 
\begin{figure*}[t!]
    \centering
    \includegraphics[width=0.7\textwidth]{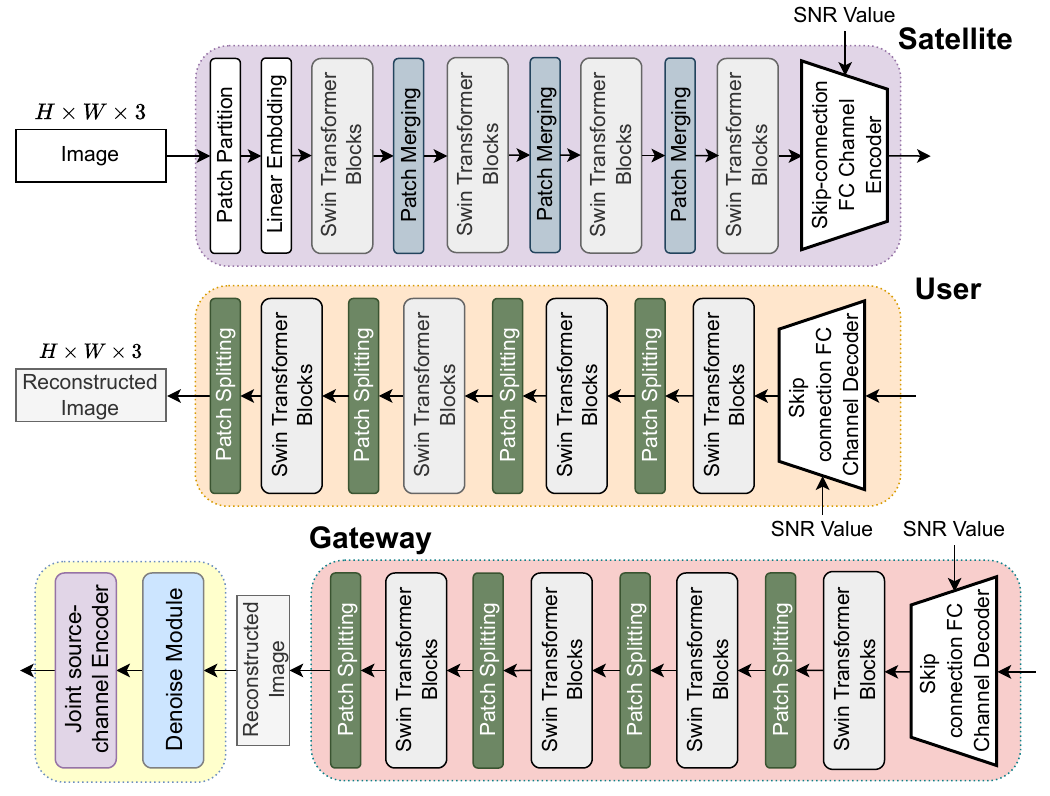}
    \caption{The architecture of the joint source-channel encoder/decoder of the satellite, GU, and the gateway.}
    \label{Sequence}
    \vspace{-0.2in}
\end{figure*}
\subsubsection{Channel Coder}
We implement a simple yet effective architecture for the channel coders to compress the extracted semantic features from the source encoder. Additionally, while SNR information cannot reveal the exact noise, it does indicate the noise range, which is extremely useful for combating environmental noise. Therefore, we incorporate it into the channel encoder network by passing it through three fully connected layers before combining it with the encoded semantic features using an element-wise operation. For a more detailed description of the channel coder, readers can refer to \cite{yang2023witt}. As shown in the Algorithm~\ref{alg:Alg1}, we first train the satellite and the gateway parameters, then train the GU parameters to adapt to the satellite network.

\subsubsection{Autoencoder-based Denoising Module} The autoencoder structure is famous for its ability to learn the latent representation, which motivates us to embrace architecture as our denoising model. In particular, the first part of the autoencoder structure is mapping the reconstructed image to a lower dimension. This process not only captures the critical feature of the image but also filters out noise, which typically contributes little to the essential structure of the image. Then, in the later part of the autoencoder model, the image is mapped to pictures with reduced noise. Besides, we acknowledged that the randomness of the noise from the environment will affect the transmitting signal in different ways under the same SNR, thus leading to different reconstructed images that even transmit the same signal. Thus, we propose a multi-view denoise technique that can mitigate this phenomenon by sampling different noised images with the same SNR value and then minimizing the averaged mean square error among those noised images toward the originally transmitted one. The process can be found in Algorithm~\ref{algorithm2}.

The denoising module empowers the satellites to effectively suppress environmental noise in signals transmitted over extended distances. A key benefit of our denoising model is its independence from end-to-end SC training. This enables its seamless deployment across various devices without the need for retraining. The proposed sequence is illustrated in Fig.~\ref{Sequence}.
\vspace{-0.2in}
\begin{algorithm}[t]
   \caption{\strut Training Process for the Proposed SC System} 
   \label{alg:Alg1}
   \begin{algorithmic}[1]
       \State{\textbf{Initialize:} Satellite Encoder ($E_{\alpha}, C_{\beta}$), Gateway Decoder ($C^{-1}_{\phi_g},D_{\theta_g}$), GU $u$ Decoder ($C^{-1}_{\phi_{u}},D_{\theta_{u}}$), training epochs for each decoder $T$.}
        \State{\textbf{Satellite \& Gateway Training}}
        \For{each epoch t$=1,2,...,T$}
          \For{each image batch}
                \State{Sampling noise $N$ based on the SNR value $\delta$} 
                \State{$X_{I}= C_{\beta}(E_{\alpha}(I),\delta)$}
                \State{$Y_{I}= X_{I}+N,$}
                \State{$\hat{I}= D_{\theta_g}(C^{-1}_{\phi_g}(Y_{I})),$}
                \State{$\mathcal{L}= \mathrm{MSE}(I,\hat{I})$}
                \State{Update $E_{\alpha}, C_{\beta}, C^{-1}_{\phi_g},D_{\theta_g}$ with $\nabla\mathcal{L}$.}
          \EndFor
                \EndFor
        \State{\textbf{Training GU parameters}: Identical to the above training process, but only GU parameters are updated by the loss.}
    \State{\textbf{Output:} Trained SC system.}
   \end{algorithmic}
\end{algorithm}
\begin{algorithm}[t]
   \caption{\strut The Sequence to Train Denoising Module} 
   \label{algorithm2}
    \textbf{Step 1:} Train all the encoders and decoders of Satellite, Gateway, and GU with Algorithm~\ref{alg:Alg1}.\\
    \textbf{Step 2}: Encode, transmit and decode an image over a noisy environment two times with the same SNR value but different noise values to create different versions of noise images.\\
    \textbf{Step 3}: Train the autoencoder-based denoising module with the multi-view images.
\end{algorithm}
\vspace{-0.2in}
\subsection{Pathways and Subcarriers Selection to Meet QoS Demand}
\subsubsection{Sub-problems}
With the trained SC system, the proposed latency problem remains dependent on the pathways of the transmissions. The problem is not only about whether direct or indirect transmission but also which subcarrier bandwidth is allocated to which GU to acquire the lowest latency. Therefore, it is a discrete optimization problem. In addition, it highly depends on GU demands for image quality and physical environment conditions, which make the situation even more challenging to resolve. The pathways and subcarriers selection problem is described as follows:
\begin{mini!}|s|[2]<b>                 
		{\substack{\boldsymbol{\gamma}, \boldsymbol{\chi}}  } 
		{ \boldsymbol{O}(\boldsymbol{\gamma}, \boldsymbol{\chi})  }{\label{P2}}{\textbf{P2:}}	
		\addConstraint{\gamma^{k}_{m} \in \{0,1\}, \quad \forall k \in \mathcal{K}, ~\forall m \in \mathcal{M}} {\label{prob_association_constraint_1b} }
		\addConstraint{\sum_{m=1}^{M} \gamma^{k}_{m} \leq 1, \quad \forall k \in \mathcal{K}} {\label{prob_association_constraint_2b}}
		\addConstraint{ \sum_{k=1}^{K} \gamma^{k}_{m} \leq 1, \quad \forall m \in \mathcal{M}} {\label{prob_association_constraint_3b}}
        \addConstraint{\chi^j_{u'} \in \{0,1\}, \quad \forall j \in \mathcal{J}, ~\forall u' \in \mathcal{U}-u} {\label{prob_association_constraint_4b} }
		\addConstraint{\sum_{u'=1}^{U-u} \chi^j_{u'} \leq 1, \quad \forall j \in \mathcal{J}} {\label{prob_association_constraint_5b}}
		\addConstraint{ \sum_{j=1}^{J} \chi^j_{u'} \leq 1, \quad \forall u' \in \mathcal{U}-u} {\label{prob_association_constraint_6b}}
		\addConstraint{\mathbb{E}(\mathrm{PSNR}_u(\boldsymbol{X}))} \geq \Psi_u, \quad \forall u \in \mathcal{U} {\label{PNSR_Constraint_7b}}
  		\addConstraint{\mathbb{E}(\mathrm{PSNR}_{u'}(\boldsymbol{X}))} \geq \Psi_{u'}, \quad \forall u' \in \mathcal{U}-u' {\label{PNSR_Constraint_8b}}
  	\addConstraint{0 \leq \gamma^{k} t_{s,m} \leq T_m, \quad \forall m \in \mathcal{M},~k \in \mathcal{K}}.    {\label{Access_window_Constraint_8b}}
\end{mini!}

The optimization of satellite-to-GU/gateway and gateway-to-bad SNR GU paths, as well as subcarrier allocation, is presented in the subsequent subsection.
\subsubsection{Discrete Whale Optimization Algorithm (DWOA) Path Selection \& Subcarrier Allocation}
We solve both problems with an effective algorithm. Specifically, we develop a two-stage DWOA to solve the pathways and subcarriers problem from the satellite to the GUs/gateway in the first stage. Then the second stage is to allocate the subcarrier frequency from the gateway to GUs. The problem formulation for satellite to gateways is given in (\ref{P2.1}):
\begin{mini!}|s|[2]<b>                 
		{\substack{\boldsymbol{\gamma}}} 
		{\boldsymbol{O}(\boldsymbol{\gamma})  }{\label{P2.1}}{\textbf{P2.1:}}	
		\addConstraint{\gamma^k_{m} \in \{0,1\}, \quad \forall k \in \mathcal{K}, ~\forall m \in \mathcal{M}} {\label{prob_association_constraint_P2.1.1} }
		\addConstraint{\sum_{m=1}^{M} \gamma^{k}_{m} \leq 1, \quad \forall k \in \mathcal{K}} {\label{prob_association_constraint_P2.1.2}}
		\addConstraint{ \sum_{k=1}^{K} \gamma^{k}_{m} \leq 1, \quad \forall m \in \mathcal{M}} {\label{prob_association_constraint_P2.1.3}}
		\addConstraint{\mathbb{E}(\mathrm{PSNR}_u(\boldsymbol{X}))} \geq \Psi_u, \quad \forall u \in \mathcal{U} {\label{PNSR_Constraint_P2.1.4}}
  	\addConstraint{0 \leq \gamma^{k} t_{s,m} \leq T_m, \quad       \forall m \in \mathcal{M}},      {\label{Access_window_Constraint_P2.1.5}}
\end{mini!}

The DWOA is a powerful meta-heuristic technique designed to efficiently solve complex optimization problems involving integer decision variables, as encountered in our satellite-ground communication system. Specifically, the path selection between satellites and ground users (GUs) influences not only the transmission latency but also the quality of the reconstructed satellite images. As shown in constraints (\ref{PNSR_Constraint_P2.1.4}), GUs have varying service quality demands, which necessitates distinct path selections for each GU. To address the challenge of integer-based decision variables in the optimization problems (\ref{P2.1}), we adopt the DWOA \cite{DWOA}. This algorithm adapts the traditional WOA to handle integer decision variables effectively. The DWOA initializes a population of candidate solutions (i.e., integer-based paths) within the feasible search space, iteratively refining these solutions through distinct exploration and exploitation phases which are defined in the subsequent sections.

\textbf {Exploitation Phase}: During the exploitation phase, DWOA mimics the bubble-net hunting strategy of humpback whales. The algorithm focuses on refining the current best solutions by adjusting them based on their proximity to other high-quality solutions. The update of the candidate solution $\Vec{\gamma}_z (\tau+1)$ as:
\begin{equation}
    \Vec{\gamma}_{z} (\tau+1)=\left \{ \begin{array}{ll}{\Vec{\gamma}_{z}^{*}(\tau)-\Vec{A}\cdot\Vec{D},} & {\textrm{p $<$0.5,}} \\ 
      \\{\Vec{D}' \cdot e^{bz} \cdot cos (2\pi z) + \Vec{\gamma}_{z}^{*}(\tau),} & {\textrm{p $\geq$0.5,}}\end{array}\right. \label{exploitation}
\end{equation}
where $ \Vec \gamma^*_z (\tau)$ denotes the best solution at iteration $\tau$, $p$ is a random variable controlling the balance between two behaviors (shrinking encircling and spiral), $\Vec{A}$ and $\Vec{D}$ are coefficient vectors used to guide the exploitation process, and $z$ represents the candidate's current position in the solution space. The first condition mimics the whale's encircling behavior, while the second accounts for the whale’s spiral movement toward prey.

\textbf {Exploration Phase}: In the exploration phase, DWOA promotes diversity in the population by encouraging candidate solutions to search the space globally. Each candidate solution moves towards a randomly selected solution in the population, allowing the algorithm to explore new areas of the search space. This can be mathematically described as:
\begin{equation}
    \Vec{\gamma}_{z}(\tau+1)=  \Vec{\gamma}_{{rand}}(\tau)-\Vec{A}\cdot\Vec{D}.  \label{exploration}
\end{equation}
where $\Vec{\gamma}_{{rand}}(\tau)$ is the position of a randomly selected solution at iteration $\tau$, and the coefficient vectors $\Vec{A}$ and $\Vec{D}$ are updated to control the step size and direction of exploration.

\textbf{Fitness Evaluation \& Constraint Handling}: At each iteration, the fitness of the candidate solutions is evaluated based on the objective function, which in our case could be a combination of minimizing transmission latency and maximizing image quality, as described by the PSNR constraints. To handle constraint violations, the algorithm incorporates a penalty function into the objective function. This ensures that infeasible solutions are penalized, guiding the search toward the feasible region of the solution space. The fitness function for the satellite-to-GU or gateway path selection and subcarrier allocation problem can be defined as follows:
\begin{equation}
    f(\boldsymbol{\gamma}) = \boldsymbol{O(\gamma)} + \nu*C,
\end{equation}
where $\nu$ defines the coefficient for violating the QoS and $C$ denotes number of violations. 
\begin{algorithm}[t]
    \caption{Discrete Whale Optimization Algorithm for SC-QoS}
    \label{alg:DWOA}
    \begin{algorithmic}[1]
        \STATE \textbf{Input:} $\mathcal{N}$: Search agents, $\mathcal{K}$, Subcarrier sets, SNR values, \textit{MaxIT}.
        \STATE Initialize whale population: $\boldsymbol{\gamma^n_k}$ (Stage 1). Calculate initial latencies and set leading agent $\gamma^*_k$, $\tau = 0$.
        \WHILE{$\tau \leq \textit{MaxIT}$}
            \FOR{$n \gets 1$ to $\mathcal{N}$}
                \STATE Update parameters: $a$, $A$, $C$, $z$, $p$.
                \IF{$p < 0.5$} 
                    \IF{$|A| < 1$}
                        \STATE Exploitation: $\boldsymbol{\gamma^k_{\text{new}}} = \boldsymbol{\gamma^*_k} - \Vec{A} \cdot \Vec{D}$ 
                    \ELSE
                        \STATE Exploration: $\boldsymbol{\gamma^k_{\text{new}}} = \boldsymbol{\gamma_{{\text{rand}}}} - \Vec{A} \cdot \Vec{D}$ 
                    \ENDIF
                \ELSE
                    \STATE Exploitation (spiral): $\boldsymbol{\gamma^k_{\text{new}}} = \Vec{D}' \cdot e^{bz} \cdot \cos (2\pi z) + \boldsymbol{\gamma^*_k}$ 
                \ENDIF
            \ENDFOR
            \STATE Update latencies and best solution $\boldsymbol{\gamma^{k*}}$ if improved.
            \STATE $\tau \gets \tau + 1$
        \ENDWHILE
        \STATE \textbf{Output:} Optimal subcarrier allocation: $\boldsymbol{\gamma^{k*}}$.
    \end{algorithmic}
\end{algorithm}

\textbf{Algorithm Complexity}: The computational complexity of the DWOA can be expressed as:
\begin{equation}
    O(MaxIT * N * (2D + f(D)))
\end{equation}
where \textit{MaxIT} denotes the maximum number of iterations, \( N \) is the population size, \( D \) represents the dimensionality of the problem (number of decision variables), and \( f(\boldsymbol{\gamma}) \) is the complexity of evaluating the fitness function. This indicates that the DWOA scales well with the problem size, making it suitable for large-scale integer optimization problems. A comprehensive overview of the solution approach is presented in Algorithm \ref{alg:DWOA}.

Similarly, the problem formulation for gateways to bad SNR GUs is given in (\ref{P2.2}):
\begin{mini!}|s|[2]<b>                
		{\substack{\boldsymbol{\chi} }  } 
		{ \boldsymbol{O}(\boldsymbol{\chi})  }{\label{P2.2}}{\textbf{P2.2:}}	
        \addConstraint{\chi^j_{u'} \in \{0,1\}, \quad \forall j \in \mathcal{J}, ~\forall u' \in \mathcal{U}-u} {\label{prob_association_constraint_P2.2.1} }
		\addConstraint{\sum_{u'=1}^{U-u} \gamma^j_{u'} \leq 1, \quad \forall j \in \mathcal{J}} {\label{prob_association_constraint_P2.2.2}}
		\addConstraint{ \sum_{j=1}^{J} \gamma^j_{u'} \leq 1, \quad \forall u' \in \mathcal{U}-u} {\label{prob_association_constraint_P2.2.3}}
		\addConstraint{\mathbb{E}(\mathrm{PSNR}_{u'}(\boldsymbol{X}))} \geq \Psi_{u'}, \quad \forall u' \in \mathcal{U}-u {\label{PNSR_Constraint_P2.2.4}}
\end{mini!}
Given the identical objective function and constraints of problems (\ref{P2.2}) and (\ref{P2.1}), the DWOA algorithm, previously employed for (\ref{P2.1}), is also applicable to (\ref{P2.2}). To avoid redundancy, the comprehensive solution algorithm is provided in the subsequent subsection.

\subsubsection{DWOA Implementation for SC-QoS, Path Selection \& Subcarrier Allocation Optimization}
We apply the DWOA in a two-stage pathway selection algorithm for subcarrier allocation and path optimization in our satellite-ground communication system. The algorithm is structured as follows:
\begin{enumerate}
    \item Stage 1: Select optimal pathways and allocate subcarrier frequencies from the satellite to ground SNR GUs or gateways using the DWOA.
    \item Stage 2: For indirect connections via gateways, the algorithm further optimizes the total transmission latency by adjusting the bandwidth allocation from the gateway to the bad SNR GUs.
\end{enumerate}
The complete procedure is described in Algorithm \ref{alg:2-Stage-DWOA}. The optimal pathways and subcarriers \( \boldsymbol{\gamma^{*}} \) and \( \boldsymbol{\chi^*} \) are determined for both direct and indirect communication links. By leveraging the discrete nature of DWOA, our approach efficiently tackles the integer constraints in path selection and resource allocation, yielding high-quality solutions for satellite-ground communication systems in 6G NTNs. 
\begin{algorithm}[t]
    \caption{2-Stage Pathway Selection Using DWOA for SC-QoS Problem}
    \label{alg:2-Stage-DWOA}
    \begin{algorithmic}[1]
        \STATE \textbf{Input:} $\Gamma$: Environment conditions, $\Psi_u, \Psi_{u'}$: QoS demands, $T_m$: access window.
        \STATE \textbf{Stage 1:} Use DWOA to select pathways and allocate subcarriers for direct satellite-to-GU/gateway communication, minimizing latency while meeting QoS.
        \STATE \textbf{Stage 2:} For multi-hop (gateway to GU), DWOA optimizes subcarrier allocation to minimize total latency and meet QoS demands.
        \STATE \textbf{Output:} Optimal pathways and subcarrier allocations: $\boldsymbol{\gamma^*}$ and $\boldsymbol{\chi^*}$ for direct and indirect links, respectively.
    \end{algorithmic}
\end{algorithm}
\section{Simulation Results \& Discussion}
\label{sim_results}
\subsection{Simulation Setup, Training \& Evaluating Dataset}
To validate the proposed gateway-assisted SC framework and assess the effectiveness of the two-stage DWOA, we conducted a series of simulations designed to highlight the advantages of our system. The simulation environment was carefully established, with benchmark schemes introduced for comparative analysis. Key parameters for the communication settings are summarized in Table~\ref{sim_tab}. We experimented with both AWGN and Rayleigh fading channels across a wide range of noise power levels. Notably, the noise power between the satellite and the joint set of $M$ devices was observed to be lower than that between the gateway and GUs, primarily due to the presence of obstacles in the latter communication link, which introduce greater interference and signal degradation. We utilize high-resolution images from the \textit{DIV2K} training dataset \cite{agustsson2017ntire} to train the SC network. The performance of the trained network is then evaluated using both the \textit{DIV2K} dataset and the \textit{FloodNet} dataset \cite{9460988} to comprehensively demonstrate the effectiveness of the proposed approach. For the training process, a learning rate of $5e^{-4}$ is employed, and the Adam optimizer is utilized to optimize the deep learning models.
\begin{table}[t]
\centering
\setlength{\arrayrulewidth}{0.10mm}
\setlength{\tabcolsep}{1pt}
\renewcommand{\arraystretch}{1}
\caption{Simulation Parameters}
\label{sim_tab}
\scalebox{1.1}{
\begin{tabular}{|c|l|c|}
\hline
    \textbf{Notation}& \textbf{Definition}& \textbf{Value} \\ \hline \hline
    $\mathcal{U}$ & Number of GUs, i.e., $u \cup u'$ & $ 20 $\\ \hline
    $f^{k}_{s}$ & Satellite subcarrier frequency  & $[20-30]$~GHz\\ \hline
    $B_{s}$ & Bandwidth of satellite subcarrier & $500$~MHz\\ \hline
    $G_s~$ & Satellite antenna gain & $33.13$~dBi \\ \hline
    $G_u~$ & GU antenna gain & $10.4$~dBi \\ \hline
    $G_g~$ & Gateway antenna gain & $59.0$~dBi \\ \hline
    $P_s~$ & Satellite transmit power & $~10$~W\\ \hline
    $d~$ & Satellite altitude & $786$~km \\ \hline
    $T_k~$ & Access window & $60$~sec\\ \hline
    $N_0~$ & Satellite \& GUs noise power & $\mathcal{N}(-44,1)$~dB\\ \hline
    $B_{g}$ & Gateway subcarrier frequency& $15-20$~GHz\\ \hline
    $P_{g}$ & Gateway transmit power & $~1$~W\\ \hline
    $N_1~$ & Gateway \& GUs noise power & $\mathcal{N}(-33,2)$~dB\\ \hline
    $d_{g}~$ & Gateway \& GUs distances & $10$~km \\ \hline
    $\mathrm{SCR}$ & Semantic compression ratio & $1/16$ \\ \hline
 \end{tabular}}
\end{table}
\subsection{Benchmarks for Performance Comparison}
To evaluate the performance improvements achieved through gateway-assisted SC and subcarrier frequency allocation in satellite networks, we consider the following benchmark schemes for comparison:
\begin{itemize}
    \item \textit{Direct Communication \& Greedy Algorithm}: In this approach, the satellite establishes direct connections with all GUs in the network, and subcarrier frequency allocation is determined using the greedy algorithm (GRE).  
    \item \textit{Direct Communication \& DWOA}: Similar to the previous benchmark, the satellite directly communicates with all GUs; however, the subcarrier frequencies are allocated using the two-stage DWOA.
    \item \textit{Hop Communication with Gateway Assisted \& Greedy Algorithm}: In this scheme, the satellite connects indirectly to GUs with low SNR through hop routing via a gateway. The allocation of subcarrier frequencies is handled by the GRE.
    \item \textit{Hop Communication with Gateway Assisted \& DWOA}: In this approach, the DWOA is employed to determine both the direct or indirect connection strategy for all GUs in the network and the corresponding subcarrier frequency allocation, optimizing network performance.
\end{itemize}
\subsection{Results Analysis}
\begin{figure}[t]
    \centering
    \includegraphics[width=0.9\columnwidth]{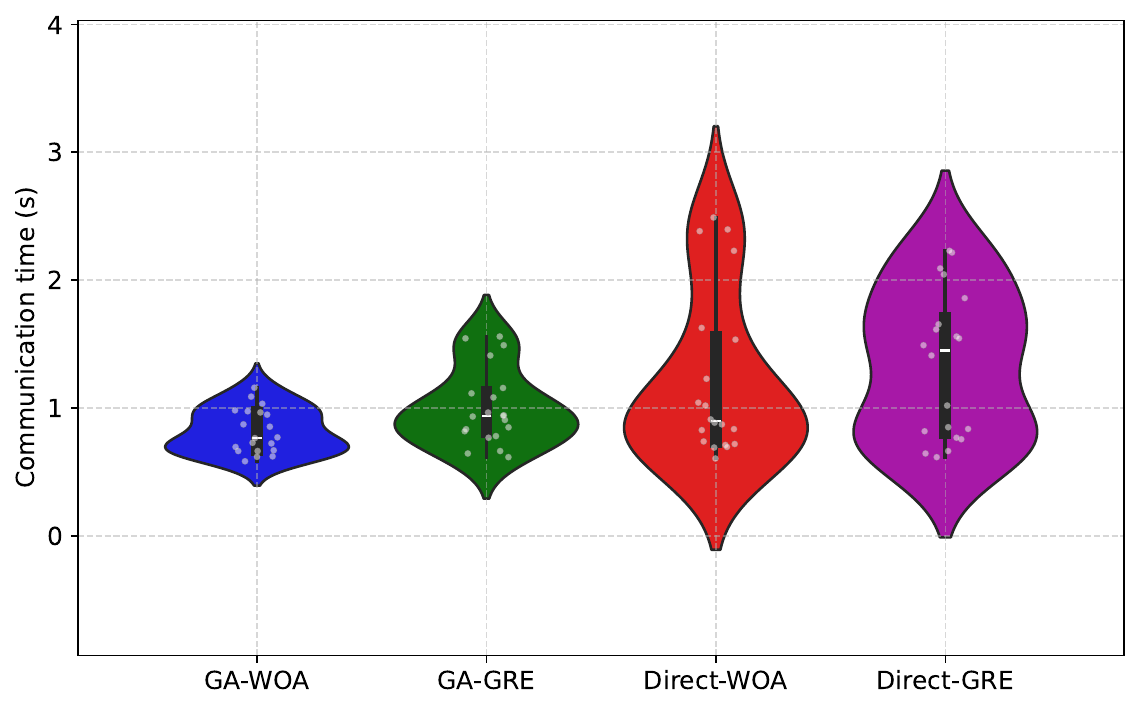}
    \caption{Comparative analysis of communication times for GUs in direct and gateway-assisted scenarios employing WOA and greedy algorithms.}
    \label{CommunicationTime}
   \vspace{-0.2in}
\end{figure}
\subsubsection{The Advantages of Gateway Assistance \& Subcarrier Allocation}
In Fig.~\ref{CommunicationTime}, we present a comparison of communication time between the direct connection scenario and the gateway-assisted approach. In the gateway-assisted configuration, the satellite establishes direct connections with half of the GUs, while the remaining half receives signals through a hop technique via the gateway. The results demonstrate that the gateway-assisted scenario significantly reduces the communication time for GUs compared to the direct connection approach. Notably, in the \textit{Direct-WOA} and \textit{Direct-GRE} scenarios, communication time varies widely among GUs due to differences in the SNR between the satellite and each GU, which heavily influences transmission time. Given the same noise power $N_{0}$, the gateway achieves a higher SNR than the GUs, benefiting from its enhanced receiving antenna gain, as shown in equation (\ref{SNRvalue}). Consequently, GUs in the gateway-assisted scenario experience more consistent and reduced communication times compared to direct communication. Furthermore, the proposed two-stage \emph{DWOA} algorithm outperforms the \emph{GRE} algorithm in terms of transmission latency, achieving an average latency of $0.8186$ seconds compared to $1.0008$ seconds with \emph{GRE}.
\begin{figure}[t]
    \centering
    \includegraphics[width=0.9\columnwidth]{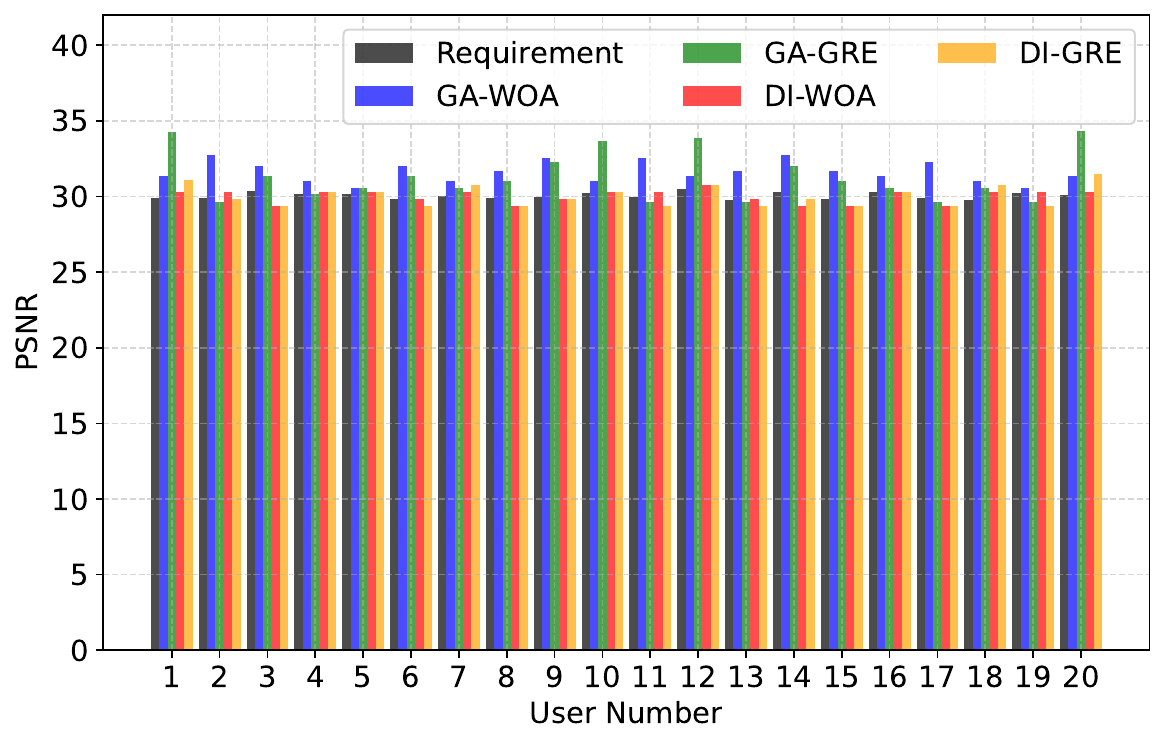}
    \caption{Required PSNR and secured PSNR under different scenarios.}
    \label{PSNRValues}
   \vspace{-0.2in}
\end{figure}
\begin{table*}[t]
\centering
\setlength{\tabcolsep}{1.5pt}
\centering
\caption{PSNR and MS-SSMI performance of GU, Gateway, Gateway with Denoise module under AWGN and Rayleigh fading channel}
\label{PSNRMSSIM}
\renewcommand{\arraystretch}{1.30}
\begin{tabular}{|c|cccccc|cccccc|}
\hline
Channel Type   & \multicolumn{6}{c|}{AWGN Channel}                                                                                                                                          & \multicolumn{6}{c|}{Rayleigh Fading Channel}                                                                                                                                      \\ \hline
Received Model & \multicolumn{2}{c|}{GU}                                   & \multicolumn{2}{c|}{Gateway}                                & \multicolumn{2}{c|}{Gateway\&Denoise}    & \multicolumn{2}{c|}{GU}                                   & \multicolumn{2}{c|}{Gateway}                                & \multicolumn{2}{c|}{Gateway\&Denoise}    \\ \hline
SNR value      & \multicolumn{1}{c|}{PSNR}    & \multicolumn{1}{c|}{MS-SSIM} & \multicolumn{1}{c|}{PSNR}    & \multicolumn{1}{c|}{MS-SSIM} & \multicolumn{1}{c|}{PSNR}    & MS-SSIM & \multicolumn{1}{c|}{PSNR}    & \multicolumn{1}{c|}{MS-SSIM} & \multicolumn{1}{c|}{PSNR}    & \multicolumn{1}{c|}{MS-SSIM} & \multicolumn{1}{c|}{PSNR}    & MS-SSIM \\ \hline
1 dB           & \multicolumn{1}{c|}{28.7946} & \multicolumn{1}{c|}{0.9271}  & \multicolumn{1}{c|}{28.8807} & \multicolumn{1}{c|}{0.9270}   & \multicolumn{1}{c|}{29.1045} & 0.9309  & \multicolumn{1}{c|}{27.4347} & \multicolumn{1}{c|}{0.9061}  & \multicolumn{1}{c|}{27.5441} & \multicolumn{1}{c|}{0.9077}  & \multicolumn{1}{c|}{27.7571} & 0.9124  \\ \hline
3 dB           & \multicolumn{1}{c|}{29.8556} & \multicolumn{1}{c|}{0.9464}  & \multicolumn{1}{c|}{29.9619} & \multicolumn{1}{c|}{0.9472}  & \multicolumn{1}{c|}{30.1458} & 0.9493  & \multicolumn{1}{c|}{28.1708} & \multicolumn{1}{c|}{0.9253}  & \multicolumn{1}{c|}{28.2585} & \multicolumn{1}{c|}{0.9266}  & \multicolumn{1}{c|}{28.4689} & 0.9300    \\ \hline
5 dB           & \multicolumn{1}{c|}{30.7371} & \multicolumn{1}{c|}{0.9588}  & \multicolumn{1}{c|}{30.8415} & \multicolumn{1}{c|}{0.9597}  & \multicolumn{1}{c|}{30.9989} & 0.9610   & \multicolumn{1}{c|}{28.7268} & \multicolumn{1}{c|}{0.9382}  & \multicolumn{1}{c|}{28.7890}  & \multicolumn{1}{c|}{0.9387}  & \multicolumn{1}{c|}{29.0160}  & 0.9415  \\ \hline
7 dB           & \multicolumn{1}{c|}{31.4642} & \multicolumn{1}{c|}{0.9672}  & \multicolumn{1}{c|}{31.5562} & \multicolumn{1}{c|}{0.9679}  & \multicolumn{1}{c|}{31.7009} & 0.9689  & \multicolumn{1}{c|}{29.1212} & \multicolumn{1}{c|}{0.9465}  & \multicolumn{1}{c|}{29.1658} & \multicolumn{1}{c|}{0.9466}  & \multicolumn{1}{c|}{29.4114} & 0.9488  \\ \hline
9 dB           & \multicolumn{1}{c|}{32.0508} & \multicolumn{1}{c|}{0.9731}  & \multicolumn{1}{c|}{32.1337} & \multicolumn{1}{c|}{0.9736}  & \multicolumn{1}{c|}{32.2722} & 0.9745  & \multicolumn{1}{c|}{29.3953} & \multicolumn{1}{c|}{0.9520}   & \multicolumn{1}{c|}{29.4291} & \multicolumn{1}{c|}{0.9518}  & \multicolumn{1}{c|}{29.7158} & 0.9545  \\ \hline
11 dB          & \multicolumn{1}{c|}{32.5138} & \multicolumn{1}{c|}{0.9774}  & \multicolumn{1}{c|}{32.5930}  & \multicolumn{1}{c|}{0.9777}  & \multicolumn{1}{c|}{32.7236} & 0.9785  & \multicolumn{1}{c|}{29.5796} & \multicolumn{1}{c|}{0.9556}  & \multicolumn{1}{c|}{29.6113} & \multicolumn{1}{c|}{0.9553}  & \multicolumn{1}{c|}{29.9221} & 0.9583  \\ \hline
13 dB          & \multicolumn{1}{c|}{32.8659} & \multicolumn{1}{c|}{0.9805}  & \multicolumn{1}{c|}{32.9466} & \multicolumn{1}{c|}{0.9807}  & \multicolumn{1}{c|}{33.0690}  & 0.9813  & \multicolumn{1}{c|}{29.7065} & \multicolumn{1}{c|}{0.9580}   & \multicolumn{1}{c|}{29.7325} & \multicolumn{1}{c|}{0.9577}  & \multicolumn{1}{c|}{30.0607} & 0.9609  \\ \hline
15 dB          & \multicolumn{1}{c|}{33.1127} & \multicolumn{1}{c|}{0.9825}  & \multicolumn{1}{c|}{33.1990}  & \multicolumn{1}{c|}{0.9827}  & \multicolumn{1}{c|}{33.3242} & 0.9833  & \multicolumn{1}{c|}{29.7919} & \multicolumn{1}{c|}{0.9594}  & \multicolumn{1}{c|}{29.8110}  & \multicolumn{1}{c|}{0.9592}  & \multicolumn{1}{c|}{30.1534} & 0.9625  \\ \hline
17 dB          & \multicolumn{1}{c|}{33.2809} & \multicolumn{1}{c|}{0.9838}  & \multicolumn{1}{c|}{33.3716} & \multicolumn{1}{c|}{0.9841}  & \multicolumn{1}{c|}{33.5036} & 0.9847  & \multicolumn{1}{c|}{29.8467} & \multicolumn{1}{c|}{0.9604}  & \multicolumn{1}{c|}{29.8625} & \multicolumn{1}{c|}{0.9605}    & \multicolumn{1}{c|}{30.2149} & 0.9635  \\ \hline
19 dB          & \multicolumn{1}{c|}{33.3895} & \multicolumn{1}{c|}{0.9846}  & \multicolumn{1}{c|}{33.4859} & \multicolumn{1}{c|}{0.9850}   & \multicolumn{1}{c|}{33.6214} & 0.9855  & \multicolumn{1}{c|}{29.8817} & \multicolumn{1}{c|}{0.9610}   & \multicolumn{1}{c|}{29.8957} & \multicolumn{1}{c|}{0.9611}  & \multicolumn{1}{c|}{30.2549} & 0.9642  \\ \hline
\end{tabular}
\end{table*}
\subsubsection{Importance of PSNR Threshold} In Fig.~\ref{PSNRValues}, we illustrate the QoS requirements for GUs and the corresponding Peak SNR (PSNR) performance across various scenarios. Each GU is assigned a unique QoS requirement for the received image, drawn from a normal distribution with a mean of $30$ and a variance of $0.2$. Among the benchmark approaches, only the proposed \emph{GA-WOA} successfully meets the QoS requirements for all GUs. In contrast, the \emph{GA-GRE} approach prioritizes minimizing transmission time for each GU but fails to account for the overall system performance, resulting in penalties for QoS violations. Specifically, we impose a $0.5$-second penalty to the communication latency for each QoS violation, representing the additional information needed to satisfy GU preferences. In scenarios where direct communication is employed between the satellite and GUs, even the \emph{DWOA} algorithm cannot meet the QoS requirements for all GUs. The subcarrier frequency allocation determined by \emph{DWOA} fails to satisfy the QoS for five GUs, while the number of QoS violations is even greater when using \emph{GRE}. These shortcomings highlight the inherent limitations of addressing the subcarrier frequency allocation problem without the benefit of gateway assistance. The results presented in the figure demonstrate the effectiveness of the proposed gateway-assisted approach and underscore the critical role of subcarrier frequency allocation in maintaining image quality for GUs.
\begin{figure}[t]
    \centering
    \includegraphics[width=0.9\columnwidth]{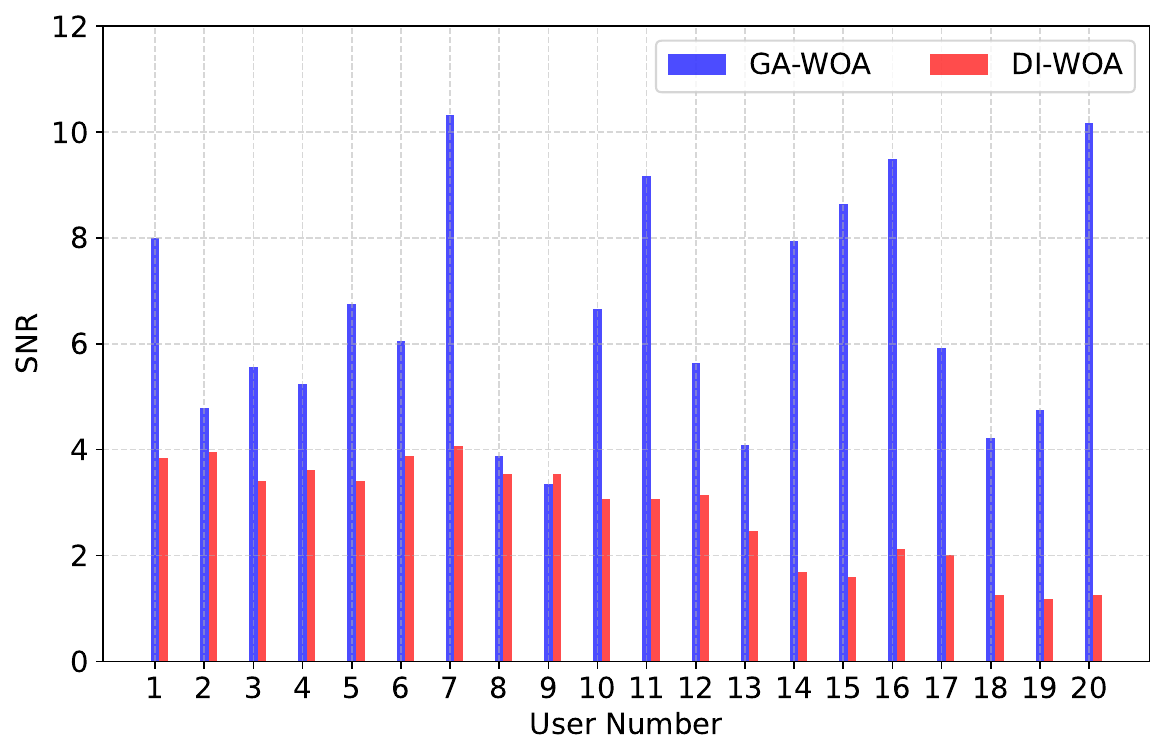}
    \caption{Each GU channel quality for successful message transmission.}
    \label{SNRValues}
   \vspace{-0.2in}
\end{figure}
\subsubsection{Impact of Channel Conditions} In Fig.~\ref{SNRValues}, we demonstrate the channel condition impact (i.e., SNR) for each GU, with subcarrier frequencies optimized by \emph{DWOA} in two scenarios: with and without gateway assistance (i.e., gateway-assisted and direct communication, respectively). Notably, GUs with low SNR values, such as GUs $14$ to $20$, are selected for hop and denoised communication via the gateway, where the higher antenna gains result in a significant SNR improvement. The indirect connection through the gateway not only boosts the SNR for GUs in high-noise environments (due to the antenna gain at the gateway) but also frees up better subcarrier frequencies for GUs with better initial conditions, allowing them to achieve even higher SNR values. For instance, GU $18$, located in an area with noise power of $6.32e^{-8}$ watts, achieves an SNR of $1.246$ with the optimized subcarrier allocation, whereas the same GU, when connected through the gateway using the same optimization algorithm, reaches an SNR of $4.222$. A similar pattern is observed with GUs $19$ and $20$, while GU $9$ experiences a slight reduction in SNR. This demonstrates the overall efficiency of the gateway-assisted approach in enhancing network performance by optimizing SNR for both low- and high-condition GUs.
\begin{figure*}[t!]
    \setcounter{figure}{7}  
    \centering
    \includegraphics[width=0.8\textwidth]{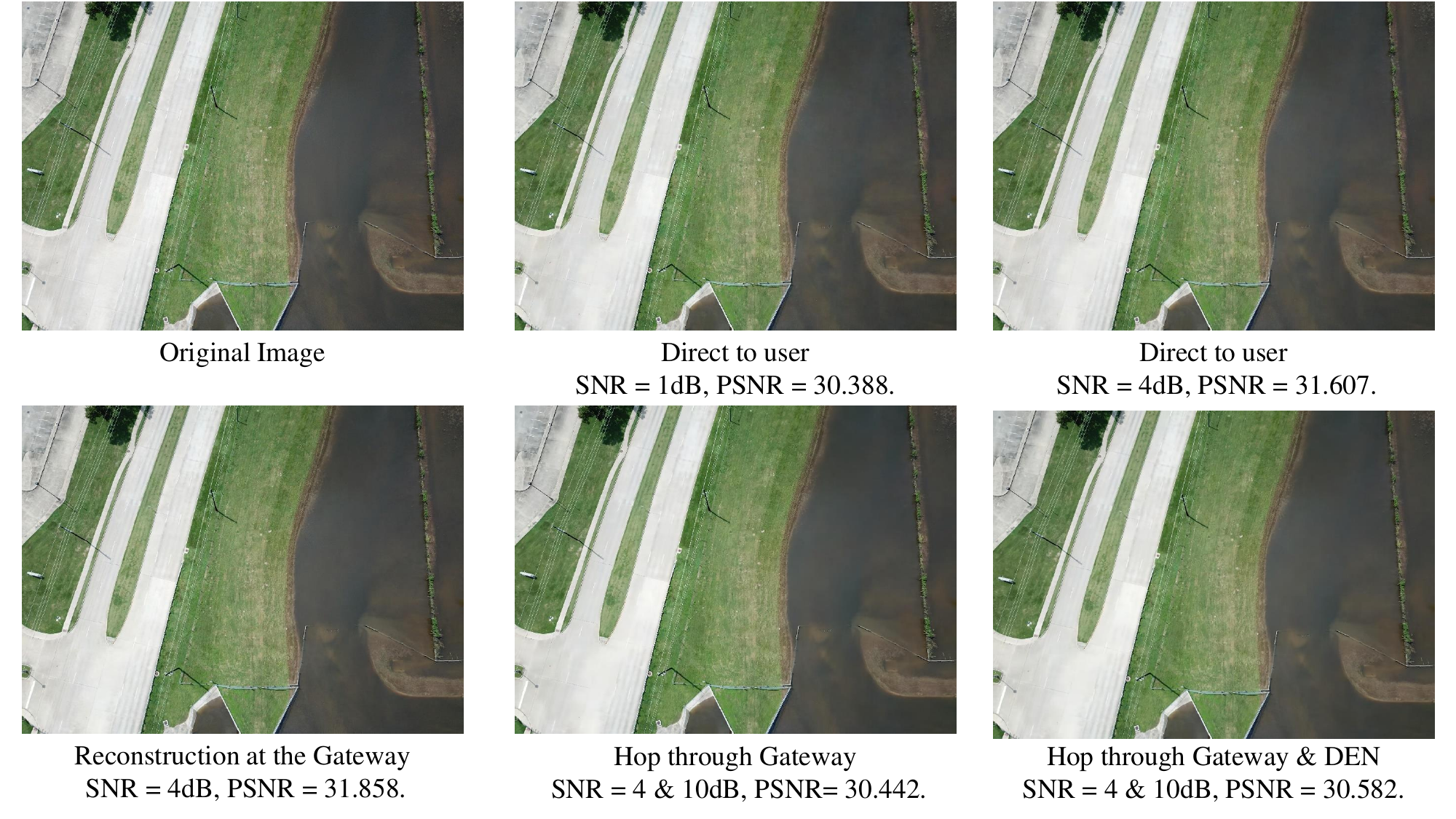}
    \caption{Reconstructed images under direct and hop connections under the AWGN channel analysis.}
    \label{Reconstructedimage}
    \vspace{-0.2in}
\end{figure*}
\begin{figure}[t]
    \setcounter{figure}{6}  
    \centering
    \includegraphics[width=0.9\columnwidth]{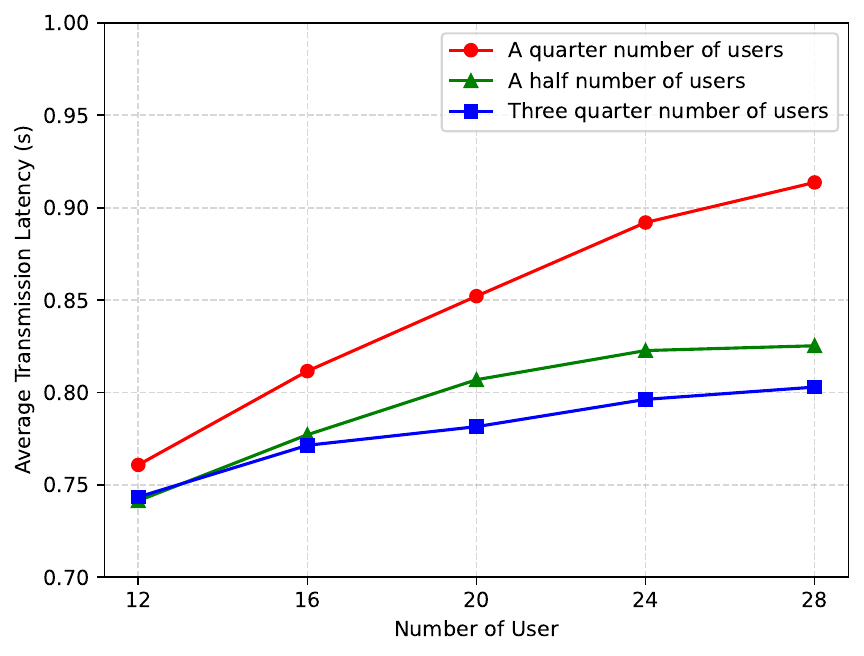}
    \caption{Average transmission latency: Impact of GU Count and assisted gateway channel number.}
    \label{Frac}
   \vspace{-0.2in}
\end{figure}

\subsubsection{The Benefits of Gateway-Assisted Communication} In our initial setup, we assumed that half of the total GUs, out of twenty, would receive assistance from the gateway. To assess the broader effectiveness of the proposed gateway assistance, we conducted experiments with varying numbers of GUs and different fractions receiving assistance: one-quarter, one-half, and three-quarters, as illustrated in Fig.~\ref{Frac}. We also adjusted the number of subcarrier frequencies to correspond with the number of GUs, while maintaining consistent bandwidth. Our findings reveal that the average transmission latency for GUs increases as the number of GUs in the network grows, across all scenarios. However, the rate of latency increase varies significantly. In the scenario where only a quarter of GUs receive gateway assistance, the latency increases rapidly. In contrast, when half or three-quarters of GUs receive assistance, the latency increase is more gradual, with the three-quarter scenario achieving the lowest overall latency. Despite this, avoiding over-relying on gateway assistance is important, as the limited availability of subcarriers between the gateway and GUs may constrain performance. Therefore, the optimal level of gateway assistance should be carefully calibrated, based on the number of GUs experiencing low SNR, to balance transmission efficiency and resource availability effectively.
\subsubsection{Advantages of the Medium-Computing Decoder \& Denoise Module at the Gateway}
In Table \ref{PSNRMSSIM}, we present the average PSNR and Multi-Scale Structural Similarity Index Measure (MS-SSIM) metrics for GUs equipped with low-computing capabilities, the gateway with medium-computing power, and the gateway equipped with a denoising module, evaluated for both AWGN and Rayleigh fading channels. Across all models, performance improves as the SNR increases, with the most rapid gains observed between $1$ and $13$ dB, followed by more gradual improvements at higher SNR levels. At any given SNR value, the low-computing GUs consistently exhibit lower performance than the medium-computing gateway, particularly at low SNR levels, with a performance gain of approximately $0.1$ in PSNR for both channels. However, the gateway equipped with the denoising module demonstrates significant performance gains, especially at lower SNR values. For instance, at 1 dB, the gateway's PSNR improves from $28.7946$ to $29.1045$ in the AWGN channel and from $27.4347$ to $27.7571$ in the Rayleigh channel. While the MS-SSIM gains of $0.0038$ and $0.0063$ may appear modest, they are substantial when considering that the maximum MS-SSIM value is $1$. These results highlight the considerable advantage of employing a denoising module at the gateway, particularly in challenging communication environments.
\subsubsection{Analysis of Image Retrieval \& Denoising} In Fig.~\ref{Reconstructedimage}, we validated the effectiveness of the proposal with the reconstruction of images at the GU through both direct connection and via gateway hop, as well as the image reconstructed at the gateway when it serves as the targeted receiver. We use the SNR values from GU $18$, as shown in Fig.~\ref{SNRValues}, to highlight the performance improvement in terms of PSNR. While the visual differences between the reconstructed images may be subtle to the human eye, PSNR provides an objective measure of image quality.
Under the same noise power, transmitting the image to the gateway instead of directly to the GU results in an SNR increase from $1$ dB to $4$ dB, significantly improving the PSNR from $30.388$ (first row, middle) to $31.858$ (second row, first). Although the noise between the gateway and the GU is greater than that between the satellite and the GU, the shorter distance between the gateway and the GU ensures a higher SNR, approximately $10$ dB. The second and last images in the second row show the result of hop transmission through the satellite gateway-GU pathway, which still outperforms direct transmission. By employing the denoising (DEN) module at the gateway, we effectively mitigate the noise effects from the previous transmission, resulting in the GU receiving a higher-quality image. This demonstrates the significant advantage of gateway-assisted hop transmission with denoising, particularly in noisy communication environments. 
\section{Conclusion}
\label{conc}
In this paper, we address the challenges of SC in satellite networks, particularly in scenarios where multiple GUs experience degraded signal quality due to satellite altitude. To enhance communication efficiency and reliability, we proposed a novel gateway-assisted framework that leverages gateway capacity to mitigate environmental noise and improve signal-to-noise ratio (SNR) for GUs. The approach involved optimizing communication overhead while ensuring data reconstruction quality, using a two-stage DWOA. Our simulation results demonstrate that the gateway-assisted strategy significantly improves image quality in high-noise environments, even with transmission distortion. Additionally, DWOA outperformed traditional greedy algorithms by reducing transmission latency while maintaining QoS for GUs. Our findings represent a key advancement in satellite-based SC networks, offering more reliable and efficient data transmission, particularly in disaster relief and time-sensitive applications. The proposed system also effectively reduces bandwidth requirements, adapts to complex satellite scenarios, and protects semantic information with acceptable transmission delays.
\bibliographystyle{IEEEtran}
\bibliography{ref}
\end{document}